\documentclass[12pt,preprint]{aastex}

\shorttitle{Photometric beaming binaries}
\shortauthors{Zucker, Mazeh \& Alexander}

\begin{document}

\title{Beaming Binaries --- a New Observational Category of
  Photometric Binary Stars}
\author{Shay Zucker\altaffilmark{1}, Tsevi Mazeh\altaffilmark{2},
and
Tal Alexander\altaffilmark{3,4}} 
\altaffiltext{1}{Dept. of Geophysics \& Planetary Sciences, 
	Raymond and Beverly Sackler Faculty of Exact Sciences, Tel
	Aviv University, Tel Aviv 69978, Israel}
\altaffiltext{2}{School of Physics \& Astronomy,
	Raymond and Beverly Sackler Faculty of Exact Sciences, Tel
	Aviv University, Tel Aviv 69978, Israel}
\altaffiltext{3}{Faculty of Physics, Weizmann Institute of Science, 
	PO Box 26, Rehovot 76100, Israel}
\altaffiltext{4}{The William Z. \& Eda Bess Novick career development chair}

\email{shayz@post.tau.ac.il,
	mazeh@post.tau.ac.il,
	Tal.Alexander@weizmann.ac.il}

\begin{abstract}
The new photometric space-borne survey missions CoRoT and Kepler will
be able to detect minute flux variations in binary stars due to
relativistic beaming caused by the line-of-sight motion of their
components. In all but very short period binaries ($P>10$\,d), these
variations will dominate over the ellipsoidal and reflection periodic
variability.  Thus, CoRoT and Kepler will discover a new observational
class: photometric beaming binary stars. We examine this new category
and the information that the photometric variations can provide. The
variations that result from the observatory heliocentric velocity can
be used to extract some spectral information even for single stars.
\end{abstract}


\section{Introduction}
\label{intro}

In 2003 \citeauthor{LoeGau2003} suggested a new {\it photometric}
method to detect extrasolar planets, based on the minute variability
of the stellar flux due to relativistic beaming induced by the star's
reflex radial velocity. \citet{RybLig1979} show that several factors
contribute to the beaming effect. The bolometric factors are the
Lorentz transformations of the radiated energy and the time intervals,
and the modification of the angular distribution of the radiated
energy (stellar aberration).  Thus, in the limit where the star's
radial (line-of-sight) velocity $v_{\mbox{\tiny R}}$ is much smaller
than the speed of light $c$, the observed bolometric flux $F$ is
modified relative to the emitted bolometric flux $F_0$ as
\begin{equation}
F=F_0\left(1+4\frac{v_{\mbox{\tiny R}}}{c}\right) \ .
\label{boldoppler}
\end{equation}
For band-pass photometry, the Doppler shift of the emitted frequency
and the spectral index of the source spectrum also have to be taken
into account, which finally yields:
\begin{equation}
F_{\nu}=F_{\nu \mbox{\tiny 0}}\left[1+(3-\alpha)\frac{v_{\mbox{\tiny R}}}{c}\right] \ ,
\label{banddoppler}
\end{equation}
where $F_{\nu}$ and $F_{\nu \mbox{\tiny 0}}$ are the observed and
emitted flux density at frequency $\nu$, respectively, and
$\alpha=d\log F/d\log\nu$ is the average spectral index around the
observed frequency. Note that Equations \ref{boldoppler} and
\ref{banddoppler} can also be obtained in a semi-classical context. In
any case, since the effect depends on the first order of
$v_{\mbox{\tiny R}}/c$ the velocities do not need to be highly
relativistic to detect the effect. Detecting beaming variability
related to orbital motion requires the very high precision provided by
photometric satellites. Even in the extremely high orbital velocities
of the stars orbiting the massive black hole in the Galactic Center,
the beaming effects cannot be measured with the low photometric
precision available for those stars \citep{Zucetal2006}.

\citet{LoeGau2003} suggested to detect stellar periodic motion
induced by an unseen planet through the beaming effect. They showed
that the amplitude of the beaming effect produced by an extrasolar
planet is of the order of micro-magnitude. Such an effect is barely
detectable by the new photometric space missions CoRoT \citep{Bag2003}
and Kepler \citep{Basetal2005}, which are aiming to find extrasolar
{\it transiting} planets, with a typical variability amplitude of
$100\,\mu$mag. The obvious advantage of \citet{LoeGau2003} approach is
its applicability to planets with almost any inclination, whereas
detecting transiting planets is limited only to planets with orbital
inclinations close to $90\degr$.

Here we examine the beaming effect in binary stars, and suggest a new
class of binaries --- beaming binaries.  These are a hybrid between
spectroscopic and ellipsoidal binaries, since the beaming binaries
will be detected by periodic photometric variations due to their
orbital radial velocity. We show that a binary with an orbital period
of $P=100\,$d has a beaming variability of at least ${\cal
O}(100\,\mu{\mathrm mag})$, which is easily detectable by CoRoT and
even more so by Kepler. We further show that for binaries with periods
longer than $10$~days, beaming variability dominates over the
competing effects of ellipsoidal and reflection variability. We
therefore expect the new satellites to harvest hundreds of previously
unknown binaries of this new class.

Beaming variability of order $100\,\mu$mag will also be induced by the
Earth's heliocentric motion relative to any observed source, single
star or binary. In spectroscopic observations, the effect of the Earth
motion is corrected in order to obtain heliocentric radial velocities.
Here we show how the dependence of this small variability on the
spectral slope at the observed bandpass can be used to probe the
spectral characteristics of single and binary stars.

\section{The amplitude of the beaming effect as compared with the
  ellipsoidal and reflection effects}
\label{compare}

The three kinds of periodic flux variations we expect to detect in
binary stars are those due to the ellipsoidal tidal deformations of
the stars, the reflection of the light of each star by its companion,
and relativistic beaming, which is the subject of this
work. \citet{LoeGau2003} have presented a rough comparison of those
three effects in the context of planet-hosting stars. We now compare
the three effects for binary stars, assuming for simplicity a circular
($e=0$) edge-on ($\sin i=90\degr$) orbit. Note that for systems with
two stellar components, the observed ellipsoidal effect is the
weighted average of the ellipsoidal effects of the two stars, whereas
the observed beaming effect is the weighted {\it difference} between
the beaming variabilities of the two stars, as their effects are
exactly in opposite phase. In this respect beaming is similar to the
reflection effect.

To a good approximation (see below), the magnitude of the beaming
effect can be calculated under the assumption that the two stars
radiate as blackbodies. For a blackbody source of temperature
$T_{\mathrm{eff}}$ the spectral index $\alpha$ is:
\begin{equation}
\alpha(\nu) = 3 - \frac{e^x}{e^x-1}x \ ,
\end{equation}
where $x = h\nu/kT_{\mathrm{eff}}$. 

The binary's orbital separation is given by Kepler's third law,
\begin{equation}
a = \left(\frac{M_1+M_2}{M_{\sun}}\right)^{1/3}
\left(\frac{P}{1\,\mathrm{year}}\right)^{2/3} \, \mathrm{AU} \ ,
\end{equation} 
where $M_i$ are the two masses, with the subscript 1 referring to the
primary and 2 to the secondary.  The amplitude of the primary's
radial-velocity variation is then
\begin{equation} 
K_1 = \left(\frac{M_2}{M_1+M_2}\right)
      \left(\frac{a}{1\,\mathrm{AU}}\right)^{-1/2}
      29.8\,\mathrm{km\,s}^{-1} \ .
\end{equation} 
A corresponding expression is obtained for the secondary by
interchanging the subscripts $1$ and $2$. The peak-to-peak amplitude
of the total expected relative flux variation from the binary due to
beaming is then
\begin{equation} 
\left( \frac{\Delta F_{\nu}}{F_{\nu}}\right)_{\mathrm{beaming}} = 
	\frac{1}{c} \frac{K_1[3-\alpha_1(\nu)]F_{\nu,\mbox{\tiny 1}} - K_2[3-\alpha_2(\nu)]F_{\nu,\mbox{\tiny 2}}}
	{F_{\nu,\mbox{\tiny 1}}+F_{\nu,\mbox{\tiny 2}}}
	\ .
\label{lumbeam}
\end{equation} 
Note that because the observed effect is the difference between the
effects of the two stars, the beaming effect vanishes for an
equal-mass binary, where the spectral characteristics are also
identical for the two components.

In order to estimate the ellipsoidal variability, we use the
expression presented by \citet{MorNaf1993} for the peak-to-peak
ellipsoidal variability of the primary:
\begin{equation} 
\left( \frac{\Delta F_{\nu,\mbox{\tiny 1}}}{F_{\nu,\mbox{\tiny 1}}}\right)_{\mathrm{ellips}} \simeq
	0.3\frac{(15+u_1)(1+\tau_1)}{3-u_1}\frac{M_2}{M_1}\left(\frac{R_1}{a}\right)^3
\label{lumellips}
\end{equation} 
Here $\tau_1$ is the gravity darkening coefficient of the primary and
$u_1$ is its limb-darkening coefficient. We calculate a similar
expression for the secondary and then weight them by the expected
blackbody fluxes in order to obtain the total relative variation of
the binary.

\citet{MorNaf1993} also provide a prescription for calculating the
amplitude of the reflection effect. They assume that each star absorbs
some of the bolometric flux of its companion, which heats the stellar
hemisphere facing the companion, inducing an asymmetric increased
emission.  Assuming blackbody radiation law,
\citeauthor{MorNaf1993} define a `luminous-efficiency' factor by:
\begin{equation}
f_{\lambda}=\left(\frac{T_2}{T_1}\right)^4 \frac{e^{x_2}-1}{e^{x_1}-1} \ ,
\label{lumeff}
\end{equation}
where $T_1$ and $T_2$ are the temperatures of the two components. Like
the beaming effect, the contributions of the reflection effects of the
two stars are in opposite phase, and the total magnitude of the effect
is the weighted difference of the two. Keeping only the leading order
terms in the radii (expressed in terms of the orbital separation), we
obtain:
\begin{equation} 
\left( \frac{\Delta F_{\nu}}{F_{\nu}} \right)_{\mathrm{reflect}} = 
	\frac{2}{3}\frac{\left(R_2/a\right)^2 f_{\lambda}^{-1} F_{\nu,\mbox{\tiny 1}} -
	                 \left(R_1/a\right)^2 f_{\lambda} F_{\nu,\mbox{\tiny 2}}}
                        {F_{\nu,\mbox{\tiny 1}}+F_{\nu,\mbox{\tiny 2}}}
\label{reflec}
\end{equation} 

Note that the ellipsoidal and reflection variabilities were calculated
assuming tidal locking and a circular orbit.  Furthermore, we use only
the leading order terms in the fractional radii. Thus, we might be
overestimating the amplitude of those effects. However, these
expressions suffice as conservative estimates for comparing with the
beaming effects.

We use Equations \ref{lumbeam}, \ref{lumellips},and \ref{reflec} to
compare the three effects for three typical binaries and for a range
of periods.  Table \ref{tabstars} presents the parameters assumed for
the stellar components.  For the purpose of this simple comparison, we
assume $u_1,u_2=0.6$ for our hypothetical stars. We calculate the
gravity darkening coefficients using the prescription in
\citet{Mor1985}.

\begin{figure}
\plotone{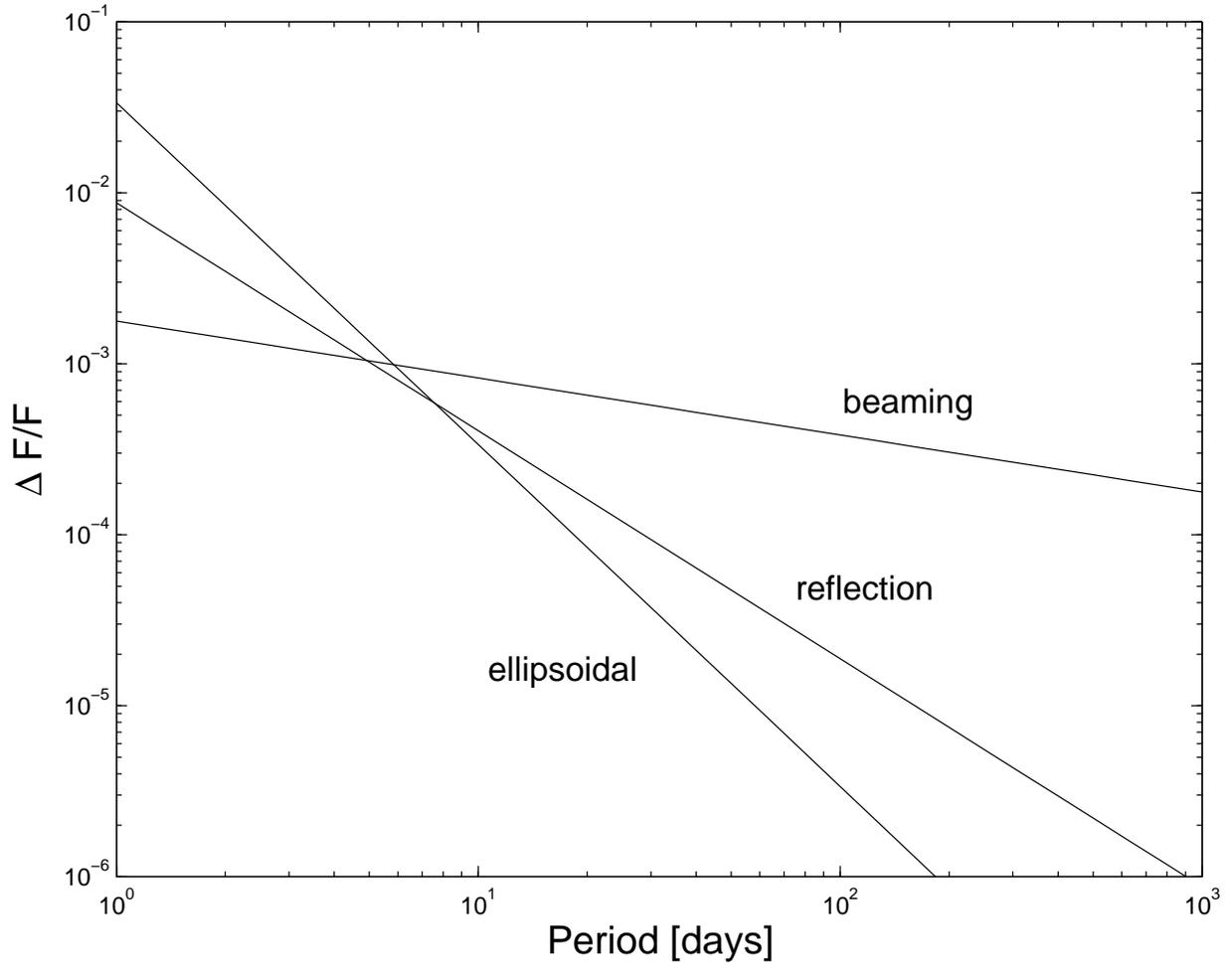}
\caption{
The three periodic photometric effects for an F0-K0 binary star in a range of periods.}
\label{compF0K0}
\end{figure}

Table \ref{tabbinaries} compares the three effects for the three
binaries, observed in the $V$ band, for periods of $10$ and $100$
days. In all cases the beaming variability dominates over the other
two effects. Figure \ref{compF0K0} shows the three effects for an
F0-K0 binary for a range of periods. The corresponding plots for the
other cases were very similar. The dependence of the effects on the
orbital separation is explicit in the expressions above, and we can
use it to understand the dependence on the orbital period. While
ellipsoidal variability decreases with period as $P^{-2}$, and the
reflection variability as $P^{-4/3}$, the beaming variability only
decreases as $P^{-1/3}$, and we expect it to become dominant for long
enough periods.

In the three cases we examined the ellipsoidal variability dominates
for periods shorter than $8$ days, while for periods longer than $10$
days the beaming variability becomes dominant.  Remarkably, the three
lines intersect at about the same period, and the reflection effect is
almost never dominant. Furthermore, the amplitude of the beaming
variability stays at the detectable levels for CoRoT and Kepler, of
$0.1-1$\,mmag for periods of a hundred days and more.

\begin{figure}
\plotone{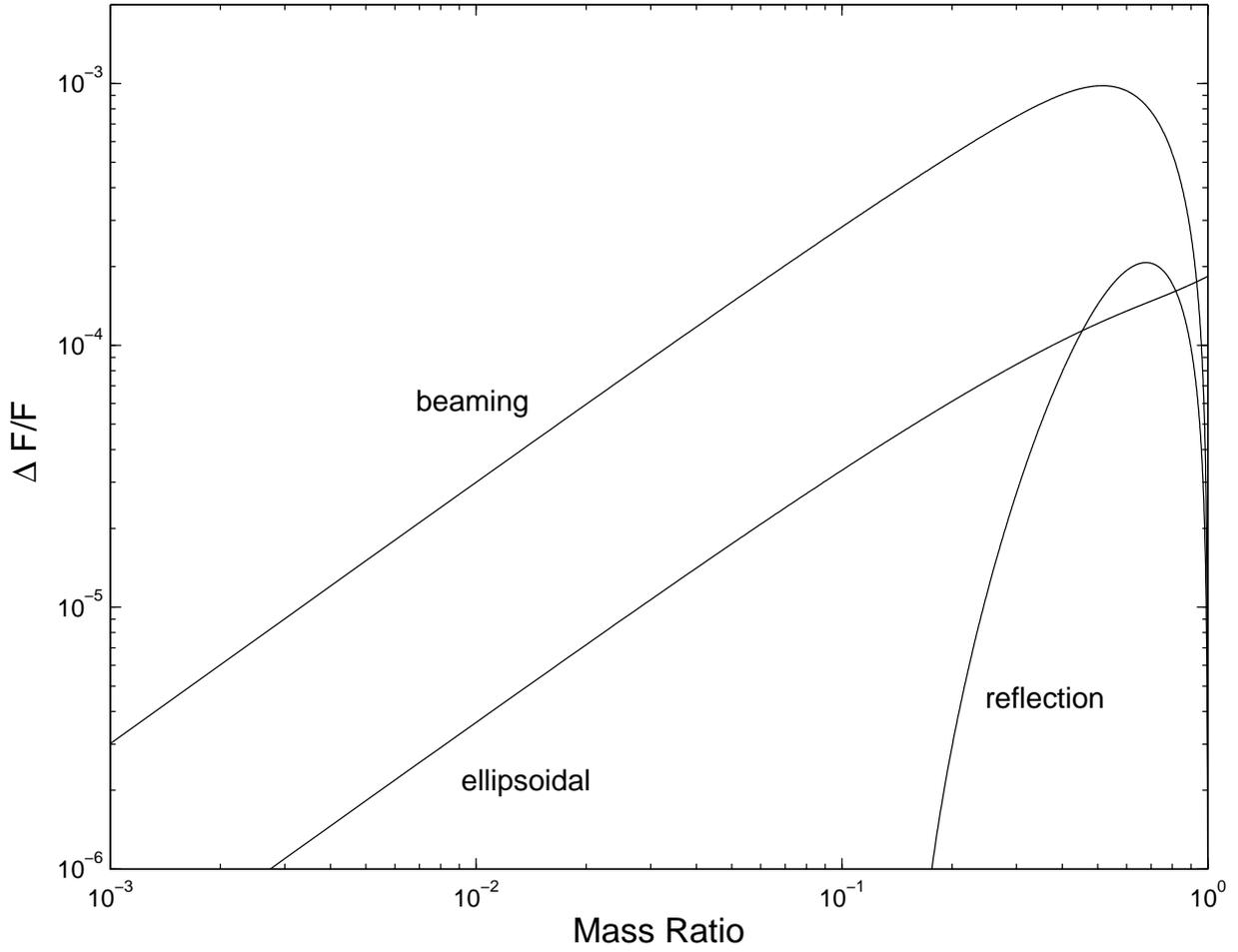}
\caption{
The three periodic photometric effects for a G0 primary in a range of
mass ratios.}
\label{figq}
\end{figure}

In Figure \ref{figq} we compare the three effects for a range of mass
ratios. We assumed a G0 primary, and used power laws for the
dependence of the secondary radius and temperature on its mass:
$R\propto M^{0.8}$ and $T_{\mathrm{eff}}\propto M^{0.55}$. While the
ellipsoidal variability is mostly sensitive to the highest mass
ratios, the beaming and reflection effects are more sensitive to
intermediary mass ratios. This is mainly because at the highest mass
ratios the effects from both binary components are canceled out.

\section{Discussion}               %
\label{discussion}                 %

The radial-velocity beaming lightcurve can yield directly most of the
spectroscopic orbital elements, including the period, eccentricity and
time of periastron passage. Since these will be obtained as the result
of a well defined magnitude-limited photometric survey, they will
provide large amounts of new data to statistical studies of
spectroscopic binaries, including, e.g., the distribution of orbital
period \citep{DuqMay1991,Mazetal2006}, and the relation between
orbital period and eccentricity \citep{Haletal2003}.

The only spectroscopic element that cannot be obtained directly from
the lightcurve is the radial-velocity amplitude $K_1$. However, as
outlined in Equation \ref{lumbeam}, the $K_1$ value can be derived
from the amplitude of the beaming effect through the spectral index
$\alpha$ of the primary and the relative amplitudes of the beaming
effect of the two components of the binary. In most binaries, the
secondary is faint enough that we will be able to ascribe the observed
beaming variability solely to the primary component. If the primary
spectral type is known, we can derive its spectral index ($\alpha$)
and obtain $K_1$, thus deriving the full set of orbital elements of a
single-lined spectroscopic binary.

In order to estimate $\alpha$ and calibrate the relation between the
beaming amplitude and $K_1$, some spectroscopic follow-up observations
of the detected binaries should be performed. Since most of the
radial-velocity elements will already be known from photometry, only a
small number of observations is needed per star.  Multi-object
spectrographs, such as FLAMES on the Very Large Telescope
\citep{Pasetal2002} or Hydra on the Wiyn telescope \citep{BarArm1995},
seem like an efficient means to obtain these observations for the
detected beaming binaries in the field.

In the few cases where the two components might have very similar
magnitudes and masses, the two contributions to the beaming
variability may cancel out, because of their opposite phases.  In
cases where the secondary light will be significant but will not
cancel the primary light completely, we will need a photometric
analogue of spectroscopic disentangling procedures like TODCOR
\citep{ZucMaz1994}. Measurements in more than one photometric band
may add the constraints needed to solve for $K_1$ and $K_2$.

We note in passing, that the derivation of $K_1$, and when possible
also $K_2$, is sensitive to any blending of the binary image with
other stars, as they depend on the {\it relative} amplitude of the
beaming effect. Therefore, it would be needed to obtain high
resolution image of the observed field, in order to spot any other
possible contributions to the binary light that might dilute the
beaming effect. In fact, measurements in different bands may also
serve the same purpose.


In addition, we propose a simple way to calibrate the relationship
between the amplitudes of the radial velocity and the beaming flux
variation. Since the satellite motion is known, and is linked with the
motion of the Earth, we already have a well known radial-velocity
signal in the data for all stars.  CoRoT, for example, will observe
dense fields around the ecliptic, continuously for almost half a
year. Thus, the amplitude of this heliocentric velocity signal will be
close to $60\,\mathrm{km\,s}^{-1}$.

\begin{figure}
\plotone{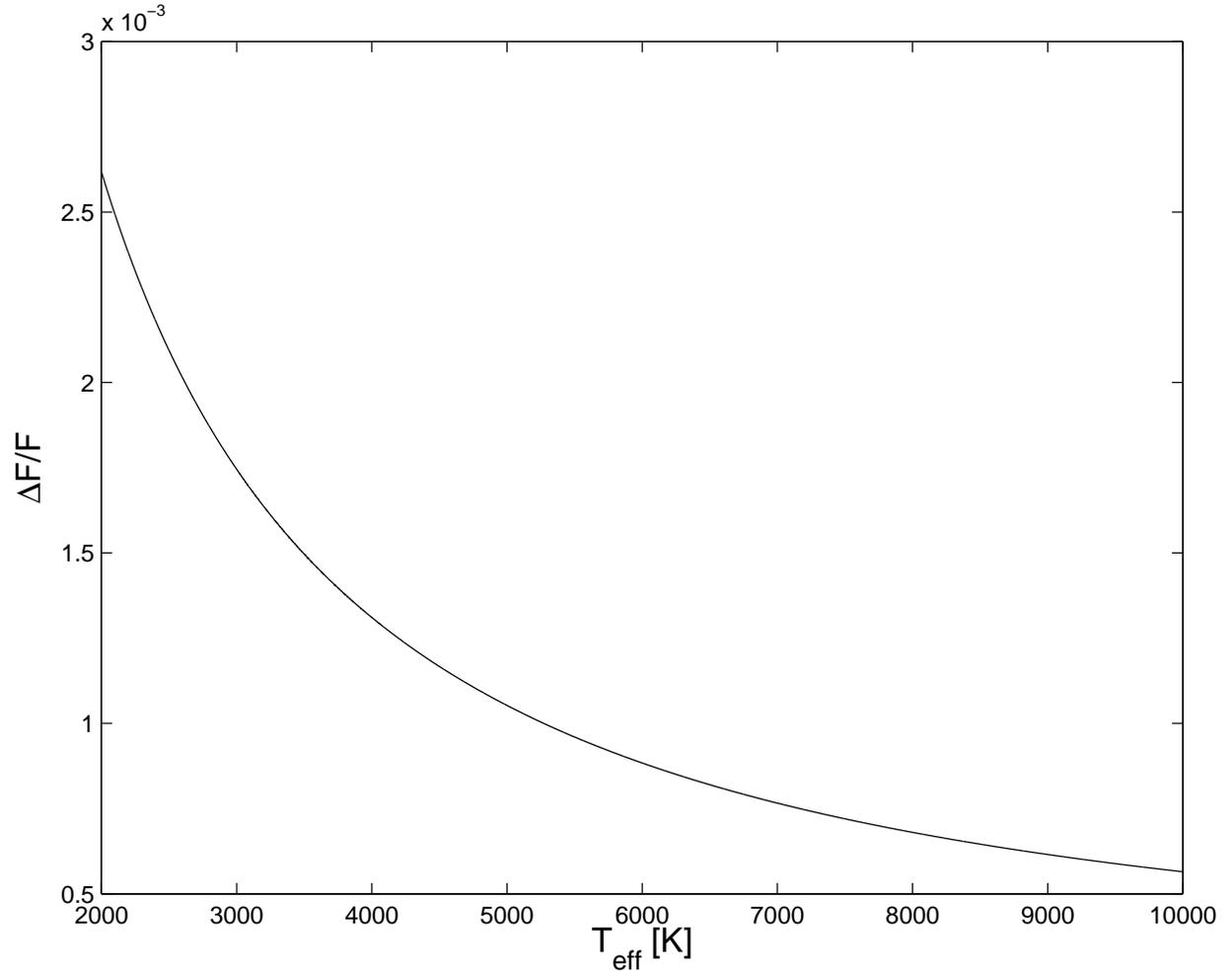}
\caption{The amplitude of the expected beaming variability of a
  black-body radiation induced by the heliocentric motion of CoRoT, in
  the V band, assuming the star location is close to the ecliptic and the 
radial velocity amplitude is $60\,\mathrm{km\,s}^{-1}$.}
\label{helio}
\end{figure}

The beaming photometric signal associated with the motion of the
telescope will affect all stars, binary and single alike.  Measuring
this signal can serve to calibrate $\alpha$ with the radial-velocity
amplitude, which in turn can be used to interpret the beaming signal
of the stellar orbital velocity, if it is a binary. For single stars
$\alpha$ is actually a piece of spectral information that reveals the
location of the passband along the blackbody radiation curve, and thus
provides an estimate of the stellar effective temperature.  Figure
\ref{helio} shows the expected photometric variability amplitude in V
for different temperatures, due to heliocentric motion alone, assuming
blackbody radiation law, and a heliocentric radial-velocity amplitude
of $60\,\mathrm{km\,s}^{-1}$.

\begin{figure}
\plotone{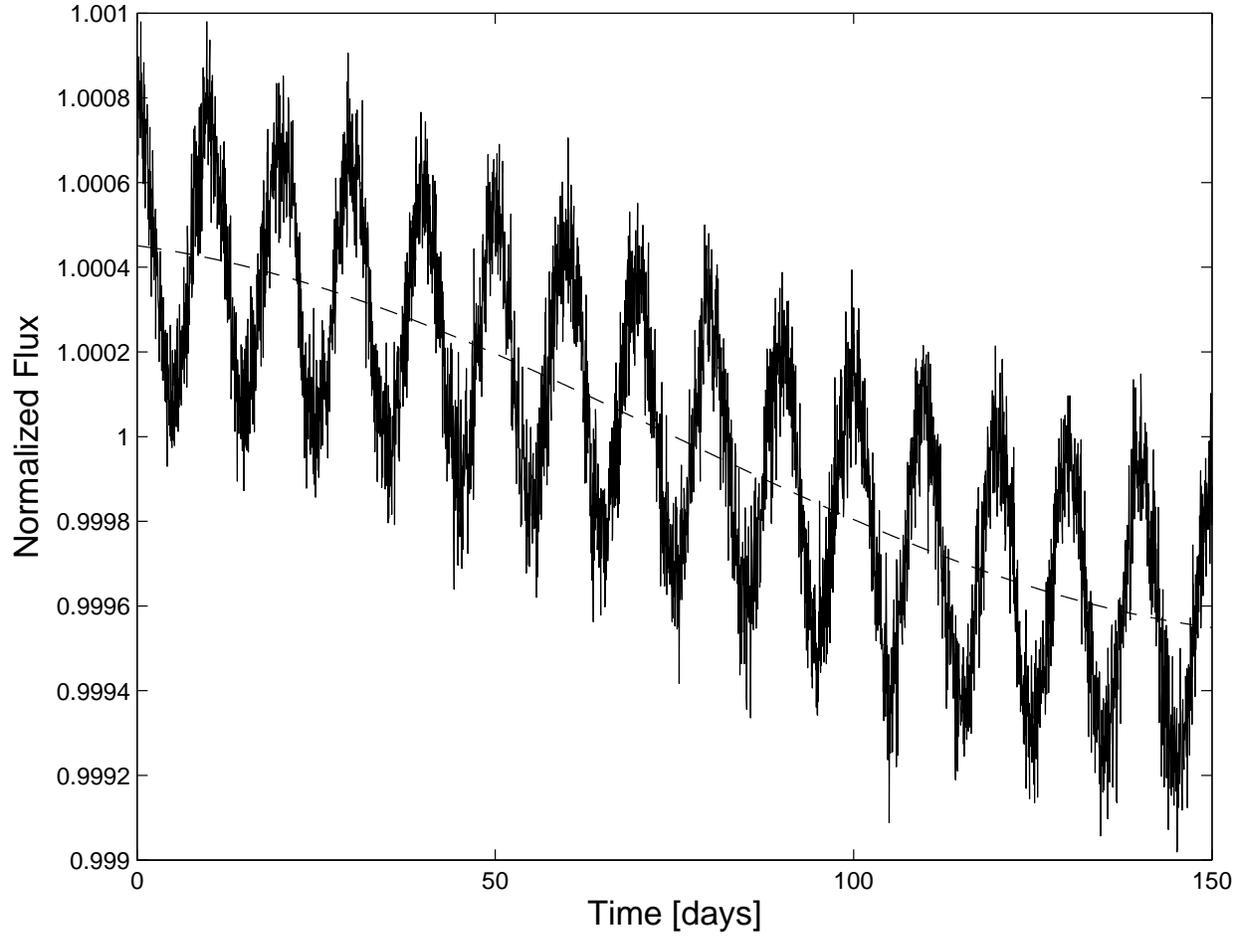}
\caption{A simulated CoRoT lightcurve of a $10$-day period G0-K0 
binary star. The beaming effects related to the binary orbit and the
heliocentric motion are easily noticed. The dashed line represents the
pure heliocentric motion beaming signal.}
\label{demolc}
\end{figure}

Figure \ref{demolc} shows a simulated lightcurve that demonstrates the
type of signal we expect to detect for a binary. The lightcurve
includes the beaming variability of a $10$-day period G0-K0 binary
star, together with the beaming variability related to the
heliocentric motion. The associated radial velocity amplitudes are
$K_1=52\,\mathrm{km\,s}^{-1}$ and $K_2=69\,\mathrm{km\,s}^{-1}$. The
noise included is only a $10^{-4}$ white noise. Note that in the case
of the heliocentric motion the contributions of the two components are
in phase, and therefore do not cancel out.

Since the signal related to the observatory motion is common to all
the observed objects, it will appear as a systematic effect, and may
be mistakenly removed as such. Algorithms such as SysRem
\citep{Tametal2005} should identify such effects and also the
individual response of each object to the same effect, and are thus
ideally suited to provide the required information for its correct
analysis.

In the lightcurve we show in Figure \ref{demolc} we have neglected
stellar microvariability. When actual lightcurves are analyzed, this
variability should be properly accounted for. For older solar-type
stars of low chromospheric activity, we may use the solar
microvariability as the only available example. Although the solar
rotational period is about $26$ days, \citet{Aigetal2004} show that
due to the short lifetimes of the spots and faculae, there is no clear
periodic signal in this period. Instead, most of the signal is in
higher frequencies, corresponding to periods shorter than $10$ days.
For non-solar type stars, variability due to chromospheric activity
may be larger than solar, and care should be taken in separating the
beaming effects from the variability effects. 

Efforts are currently underway to characterize the microvariability
that we expect to observe with the high-precision photometric
satellites \citep[e.g.,][]{Aigetal2004,Lanetal2006,Lud2006}, in order
to facilitate the detection of planetary transits. However, the
stellar microvariability is not expected to be {\it strictly}
periodic, and therefore it should be possible to single out the
beaming effect, specially because the shape of the beaming modulation
is known and depends only on a few parameters. Spectroscopic
information obtained as part of the follow-up observations should also
help to further study the chromospheric activity of the binary
candidates.

In order to give an order-of-magnitude estimate of the expected number
of beaming binaries that will be detected by, e.g., CoRoT, we estimate
that $10\%$ of the observed late-types stars have been discovered to
be spectroscopic binaries with a threshold of $K_1>3\, \mathrm{km\,
s}^{-1}$ and periods of less than a year or so.  This estimate is
based on results of the seminal works of \citet{DuqMay1991} and
\citet{Latetal2002}. During the lifetime of the mission, CoRoT is
expected to monitor about $60000$ stars. Thus, we roughly expect
$6000$ of them to be detectable as beaming binaries with $10\mu$mag
variability or higher.  Since this estimate applies to binaries with
late-type primaries, and accounting for the fact that beaming is
biased against equal-mass binaries, we can somewhat scale this number
downwards, and estimate CoRoT to yield at least $1000$ beaming binary
stars. The fact that the CoRoT sample of beaming binaries will be
discovered by a systematic, magnitude-limited survey, will augment
significantly the statistical knowledge of binaries.

\section{Conclusion}      %
As we have shown in Section \ref{compare}, we expect the multitude of
very precise light curves of CoRoT and Kepler to yield hundreds of new
binaries through their {\it periodic} beaming variability. Thus a new
observational category will emerge --- beaming binaries. In all types
of binaries, the discovery and the analysis of the binary motion
strongly depend on the timing and the number of the measurements.  For
the satellite photometric data we expect continuous radial-velocity
data that will yield the spectroscopic orbital elements, the period
and eccentricity in particular. Once the spectral index $\alpha$ is
known, the radial-velocity amplitude can be derived as well.

Without the beaming binaries, CoRoT and Kepler are supposed to find
most of the eclipsing binaries.  However, binaries with periods longer
than a few days need very fortuitous geometrical situations to present
eclipses, and are therefore rare in the data of photometric surveys
\citep[e.g.,][]{Mazetal2006}. The beaming binaries can be detected up
to a hundred days and more. Therefore, the new class of binaries will
extend our detailed knowledge of the statistical characteristics of
binaries by an order of magnitude, specially because these binaries
will emerge in the context of a well defined, complete,
magnitude-limited, photometric search. This can shed light on the
distribution of orbital period \citep[e.g.,][]{DuqMay1991,Mazetal2006}
and the relation between orbital period and eccentricity
\citep[e.g.,][]{Haletal2003}.

\acknowledgments 
This research was supported by a Grant from the G.I.F., the
German-Israeli Foundation for Scientific Research and Development to
T.M.  T.A. is supported by Minerva grant 8563 and a New Faculty grant
by Sir H. Djangoly, CBE, of London, UK.

\clearpage

\begin{deluxetable}{lcccc}
\tablecaption{Stellar parameters for simulation\label{tabstars}}
\tablehead{\colhead{parameter} & \colhead{F0} & \colhead{G0} & \colhead{K0} & \colhead{Reference}}
\startdata
Mass $[M_{\sun}]$ & 1.6 & 1.05 & 0.79 & 1 \\
Radius $[R_{\sun}]$ & 1.5 & 1.1 & 0.85 & 1 \\
Effective temperature $[\mathrm{K}]$ & 7300 & 5940 & 5150 & 1  \\
Gravity darkening coefficient in V & 0.9 & 0.4 & 0.4 & 2 \\
\enddata
\tablerefs{(1) \citet{Cox2000} (2) \citet{Mor1985}}
\end{deluxetable}

\clearpage

\begin{deluxetable}{cccccccc}
\tablecaption{The three periodic photometric effects for sample binary configurations\label{tabbinaries}}
\tablehead{\colhead{\null} & \colhead{\null} & \multicolumn{3}{c}{$P=10$~days} & \multicolumn{3}{c}{$P=100$~days} \\
\colhead{Primary} & \colhead{Secondary} & \colhead{ellipsoidal} & \colhead{reflection} & \colhead{beaming} & \colhead{ellipsoidal} & \colhead{reflection} & \colhead{beaming}}
\startdata
F0 & G0 & $3.9\cdot10^{-4}$ & $4.8\cdot10^{-4}$ & $6.4\cdot10^{-4}$ &
$3.9\cdot10^{-6}$ & $2.2\cdot10^{-5}$ & $2.9\cdot10^{-4}$ \\ F0 & K0 &
$3.4\cdot10^{-4}$ & $4.1\cdot10^{-4}$ & $8.3\cdot10^{-4}$ &
$3.4\cdot10^{-6}$ & $1.9\cdot10^{-5}$ & $3.8\cdot10^{-4}$ \\ G0 & K0 &
$1.9\cdot10^{-4}$ & $2.1\cdot10^{-4}$ & $6.6\cdot10^{-4}$ &
$1.9\cdot10^{-6}$ & $9.6\cdot10^{-6}$ & $3.1\cdot10^{-4}$
\enddata
\end{deluxetable}

\clearpage

\clearpage

\clearpage

\clearpage

\begin{thebibliography}{}

\bibitem[Aigrain, Favata \& Gilmore(2004)]{Aigetal2004}
	Aigrain, S., Favata, F., \& Gilmore, G. 2004, \aap, 414, 1139

\bibitem[Baglin(2003)]{Bag2003}
	Baglin, A. 2003, Adv. Space Res., 31, 345

\bibitem[Barden \& Armandroff(1995)]{BarArm1995}
	Barden, S. C., \& Armandroff, T. 1995, Proc. SPIE, 2476, 56

\bibitem[Basri, Borucki \& Koch(2005)]{Basetal2005}
	Basri, G., Borucki, W. J., \& Koch, D. 2005, \nar, 49, 478

\bibitem[Cox(2000)]{Cox2000}
	Cox, A. N. 2000, Allen's Astrophysical Quantities (New York:
	AIP)

\bibitem[Duqennoy \& Mayor(1991)]{DuqMay1991}
	Duquennoy, A., \& Mayor, M. 1991, \aap, 248, 485

\bibitem[Halbwachs et al.(2003)]{Haletal2003}
	Halbwachs, J. L., Mayor, M., Udry, S., \& Arenou, F. 2003,
	\aap, 397, 159

\bibitem[Lanza et al.(2006)]{Lanetal2006}
	Lanza, A. F., Messina, S., Pagano, I., \& Rodon\`{o}, M. 2006, AN, 327, 21

\bibitem[Latham et al.(2002)]{Latetal2002}
	Latham, D. W., Stefanik, R. P., Torres, G., Davis, R. J.,
	Mazeh, T., Carney, B. W., Laird, J. B., \& Morse, J. A. 2002,
	\aj, 124, 1144

\bibitem[Loeb \& Gaudi(2003)]{LoeGau2003}
	Loeb, A., \& Gaudi, B. S., \apjl, 588, L117

\bibitem[Ludwig(2006)]{Lud2006}
	Ludwig, H.-G. 2006, \aap, 445, 661

\bibitem[Mazeh, Tamuz \& North(2006)]{Mazetal2006}
	Mazeh, T., Tamuz, O., \& North, P. 2006, \mnras, 367, 1531

\bibitem[Morris(1985)]{Mor1985}
	Morris, S. L. 1985, \apj, 295, 143

\bibitem[Morris \& Naftilan(1993)]{MorNaf1993}
	Morris, S. L., \& Naftilan, S. A. 1993, \apj, 419, 344

\bibitem[Pasquini et al.(2002)]{Pasetal2002}
	Pasquini, L., et al. 2002, Messenger, 110, 1


\bibitem[Rybicki \& Lightman(1979)]{RybLig1979}
	Rybicki, G. B., \& Lightman, A. P. 1979, Radiative Processes
	in Astrophysics (New York: Wiley)


\bibitem[Tamuz, Mazeh \& Zucker(2005)]{Tametal2005}
	Tamuz, O., Mazeh, T., \& Zucker, S. 2005, \mnras, 356, 1466

\bibitem[Zucker et al.(2006)]{Zucetal2006}
	Zucker, S., Alexander, T., Gillessen, S., Eisenhauer, F., \&
	Genzel, R. 2006, \apj, 639, L21

\bibitem[Zucker \& Mazeh(1994)]{ZucMaz1994}
	Zucker, S., \& Mazeh, T. 1994, \apj, 420, 806


\end{thebibliography}
\end{document}